\documentclass[12pt,onecolumn,floatfix,superscriptaddress,tightenlines,nofootinbib]{revtex4-1}
\pdfoutput=1
\usepackage[colorlinks=true,citecolor=blue,linkcolor=blue,breaklinks=true]{hyperref}
\usepackage{amsmath,amssymb}
\usepackage{epsfig} 
\usepackage{graphicx}        
\usepackage{url}
\usepackage{color}
\usepackage{multirow}
\usepackage{placeins}
\usepackage[dvipsnames]{xcolor}
\usepackage{braket}
\usepackage{float}
\usepackage{slashed}
\hypersetup{
  colorlinks=true,
  linkcolor=red,
  filecolor=magenta,   
  urlcolor=blue,
  citecolor=blue,
} 
\providecommand{\xlink}[1]
 {\href{http://arxiv.org/abs/#1}{arXiv:#1}}
 
\allowdisplaybreaks

\bibpunct{[}{]}{,}{n}{}{,}

\def\beq{\begin{equation}}
\def\eeq{\end{equation}}
\def\bea{\begin{eqnarray}}
\def\eea{\end{eqnarray}}
\makeatletter
\@addtoreset{equation}{section}
\renewcommand{\theequation}{\thesection.\arabic{equation}}

\def\thesection{\Roman{section}}

\begin{document}

%\title{Probing Type-I seesaw with curvature-dependent lepton chemical potential}
%\title{Flavour effects in gravitational leptogenesis}
%\title{Importance of flavour effects in right  handed neutrino induced gravitational leptogenesis}
\title{Gravitational wave complementarity and impact of NANOGrav data on gravitational leptogenesis}
\author{Rome Samanta}
\email{romesamanta@gmail.com}
\affiliation{Physics and Astronomy, University of Southampton, Southampton, SO17 1BJ, U.K.}
\affiliation{CEICO, Institute of Physics of the Czech Academy of Sciences, Na Slovance 1999/2, 182 21 Prague 8, Czech Republic}
\author{Satyabrata Datta}
\email{satyabrata.datta@saha.ac.in}
\affiliation{Saha Institute of Nuclear Physics, HBNI, 1/AF Bidhannagar,
Kolkata 700064, India}

%\preprint{NUHEP-TH/19-08}

\begin{abstract} 
In seesaw mechanism, if right handed (RH) neutrino masses are generated dynamically by a gauged  $U(1)$ symmetry breaking, a stochastic gravitational wave  background (SGWB) sourced by a cosmic string network could be a potential probe of leptogenesis. We show that the leptogenesis mechanism that facilitates the dominant production of lepton asymmetry via the quantum effects of right-handed neutrinos in gravitational background, can be probed by  GW detectors as well as next-generation neutrinoless double beta decay ($0\nu\beta\beta$) experiments in a complementary way. We infer that for a successful leptogenesis,  an exclusion limit on  $f-\Omega_{\rm GW}h^2$ plane would correspond to an exclusion on the $|m_{\beta\beta}|-m_1$ plane as well. We consider a normal light neutrino mass ordering and discuss how recent NANOGrav pulsar timing data (if interpreted as GW signal) e.g., at 95$\%$ CL, would correlate with the potential discovery or null signal in $0\nu\beta\beta$ decay experiments.  
%Nevertheless, both the scenarios require an active-sterile mixing bounded from above which is beyond the reach of projected sensitivities of future collider experiments. Thus at this moment, one has to rely on the falsifiability of the mechanism upon finding a positive signal in future colliders or to look for an indirect test, e.g., in  long baseline experiments, by reducing the number of relevant model-parameters.

\end{abstract}

\maketitle

\section{Introduction}
Dominance of matter over antimatter remains one of the prominent cosmological puzzles\cite{planck,Akrami:2018vks} which needs to be addressed in a beyond Standard Model (BSM) framework. Light neutrino masses and mixing\cite{pdg,t2k,t2k1,t2k2,nova1,nova2,minos,dayabay,reno,dc,ic,sk0,globalfit}, another BSM  phenomena, can be naturally connected with matter anti-matter asymmetry with the inclusion of heavy right handed (RH) sterile states ($N_i$) that facilitate light neutrino masses via Type-I seesaw mechanism as well as lead to  lepton number violation in the theory. The simplest leptogenesis mechanism is the CP violating and out of equilibrium\cite{sk} decays of the sterile states\cite{lep1,lep2,lep3,lep4,lep5,lep6}  accompanied by a B-L conserving Sphaleron transition\cite{sp1,sp2}. There could be other sources of lepton number violation in the early universe (EU), e.g., lepton asymmetry sourced by chiral Gravitational Waves (GWs)\cite{gravlep1,gravlep2,Caldwell:2017chz,gravlep3,Papageorgiou:2017yup,Kamada,gravlep4} and by the interaction of lepton or baryon current with background gravity through the operator $\partial_\mu R j^\mu/M^2$, where $R$ is the Ricci scalar, by  the means of a dynamical CPT violation \cite{rg1,rg2,tra,Li:2004hh,Feng:2004mq,Lambiase:2006md,rg3,rg4,Lambiase:2013haa,rg5,rg6}. Interestingly, the operator  $\partial_\mu R j^\mu/M^2$ can be generated in Type-I seesaw  at two-loop level\cite{rg7,rg8,rg9,rg10} (cf. Fig.\ref{fig1}) causing a chemical potential and hence a net lepton asymmetry in equilibrium proportional to the time derivative of $R$. When it comes to the testability of standard leptogenesis within Type-I seesaw, one has either to lower the RH mass scale for collider searches\cite{Ars,Ham, dev1,dev2} or to impose restrictions on the parameter space, for example, considering discrete symmetries\cite{fasy1,fasy2,fasy3,fasy4} and theories like SO(10) grand unification (GUT)\cite{so1,so2,so3,so4,so5}. However, the discovery of GWs by LIGO and Virgo collaboration \cite{gw1,gw2,gw3,gw4,gw5,gw6,gw7} of black holes and neutron stars has opened up a new cosmic frontier for multi-frequency study of stochastic GW background (SGWB)\cite{sg1,sg2,sg3,sg4} by which many BSM theories including leptogenesis can be probed. A natural and an exciting prediction of a BSM phase transition\cite{graham1} associated with a spontaneous breaking of an Abelian symmetry is cosmic strings\cite{cs1,cs2,Nielsen:1973cs} which can form closed loop and shrink via emission of GWs\cite{Vachaspati:1984gt}. Whilst emission of GWs from cosmic string remains controversial,  numerical simulations  based on the Nambu–Goto action\cite{ng1,ng2} indicate  that cosmic string loops loose energy dominantly via GW radiation, should the underlying broken symmetry correspond to a local gauge symmetry. \\

    Most distinguishable feature of GW emission from cosmic string is prediction of a strong signal across a wide range of frequency which has triggered a growing interest in this field\cite{gr1,gr2,gr3,gr4,gr5,gr6,gr7,gr8,gr9,gr10,gr11}. This also includes recent studies to probe GUT\cite{gcs1,gcs2}, high scale leptogenesis\cite{lepcs1}, low scale leptogenesis\cite{lepcs2}. Unequivocally, GW probe of BSM models has become more interesting after the new NANOGrav analysis of 12.5 yrs pulsar timing data\cite{nanog} which reports a strong evidence for a stochastic common-spectrum process and may be interpreted as a GW signal at frequency $f\sim 1/yr$. The new data is better fitted with cosmic string models\cite{nan1,nan2,nan3} than the single value power spectral density as suggested by the models of supermassive black hole (SMBHs). For the other interpretations of the NANOGrav data, e.g., primordial black holes, dark phase transition and inflation please see Ref.\cite{nanth1,nanth2,nanth3,nanth4,nanth5,nanth6,nanth7,nanth8,nanth9,nanth10,nanth11}. \\
    
In the context of seesaw models and leptogenesis, one has a  natural motivation for a spontaneous breaking of $U(1)_{B-L}$\cite{moha1,moha2} which generates heavy RH neutrino masses as well as gives a detectable cosmic string induced GW signal.  This has been the central point of the studies in the Refs.\cite{lepcs1,lepcs2}. We go in the same direction but consider RH neutrino induced gravitational leptogenesis mechanism (RIGL)\cite{rg10} wherein lepton asymmetry is generated at two loop level due to the interactions of RH neutrinos with background gravity. A dynamical CPT violation in this process induces a lepton asymmetry in equilibrium which is maintained during the course of evolution until $\Delta L=2$ $N_1$-interaction rates fall below the Hubble expansion rate. While even without flavour effects\cite{fl1,fl2,fl3,fl4,fl5,fl6,Samanta:2018efa} in the washout processes,  the mechanism is able to produce dominant lepton asymmetry (compared to the leptogenesis from decays) \cite{rg7,rg8,rg9,rg10}, when  the effects are taken into account, the lightest RH mass scale $M_1$ can be lowered to $M_{\rm min}\sim 10^7$ GeV\cite{Samanta:2020tcl} unlike the standard $N_1$-thermal\cite{thr1,thr2,thr3,thr4} leptogenesis scenario ($T_{\rm RH}>M_1$) where it is subjected to a lower bound of $10^9$ GeV\cite{ibarra}. We consider a hierarchical spectrum of RH neutrinos assuming the heaviest mass scale is of the order of the $U(1)$ breaking scale ($\Lambda_{\rm CS}$) which also sets the initial temperature of asymmetry generation and masses of the other two are parametrically suppressed with the lightest being $\mathcal{O}(M_{\rm min})$. The magnitude of the final frozen out asymmetry depends on the heaviest mass scale as well as the strength of the $\Delta L=2$ $N_1$-interactions which typically increase with the increase of the lightest light neutrino mass $m_1$ and cause a reduction in the magnitude of the final asymmetry. This opens up the possibility to probe the RIGL mechanism in GW detectors as well as absolute neutrino mass scale experiments and consequently neutrinoless double beta decay ($0\nu\beta\beta$) experiments. We show that a successful leptogenesis corresponds to $G\mu >4.4\times 10^{-11}$  with a corresponding upper bound  $m_1 \lesssim 12$ meV, where $G$ is the Newton's constant and $\mu\sim \Lambda_{\rm CS}^2$ is the string tension. An increase in $G\mu$ causes an increase in the upper bound on $m_1$  and hence less exclusion in the $0\nu\beta\beta$ decay parameter space.  We then discuss the compatibility of RIGL mechanism with recent NANOGrav data and find that  $m_1\gtrsim 25$ meV is disfavoured by NANOGrav at 2$\sigma$.\\
    
    The rest of the paper is organised as follows: In sec.\ref{s2} we discuss the RIGL mechanism. In sec.\ref{s3} we briefly outline the production of GWs from cosmic string. In sec. \ref{s4} we present the numerical analysis and an overall discussion. 
    We summarise in sec.\ref{s5}.
\section{Right handed neutrino induced  gravitational leptogenesis}\label{s2}
\begin{figure}
\includegraphics[scale=.4]{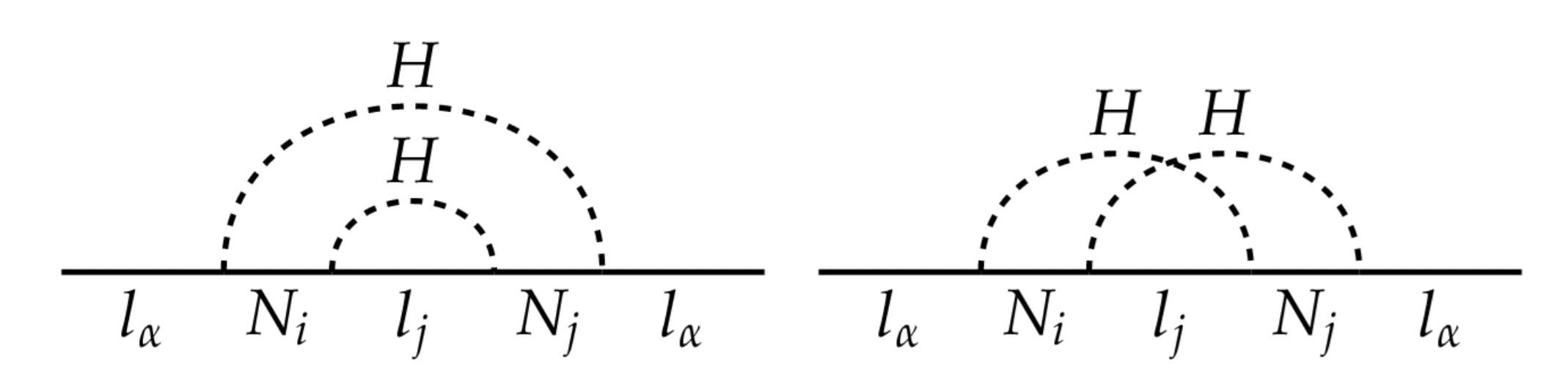}
\caption{Flat space two loop diagrams in seesaw model that generate the $\partial_\mu R j^\mu/M^2$ operator when computed in the gravitational bckground. E.g., see Ref.\cite{rg9}.}\label{fig1}
\end{figure}

The basic idea behind the gravitational leptogenesis\cite{tra} is that the C and CP violating operator  $\mathcal{L}_{CPV}\sim b\partial_\mu R j^\mu\sim b\partial_\mu R\bar{\ell}\gamma^\mu\ell$ with $b$ as a real effective coupling of mass dimension minus two, corresponds to a chemical potential $\mu=b\dot{R}$ for the lepton number in the theory.  Consequently, the normalised (by photon density $n_\gamma\sim T^3$) equilibrium lepton asymmetry which arises due to this chemical potential is given by $N_{B-L}^{eq}\sim \frac{b\dot{R}}{T}$. Now an important question that can be addressed is, that without introducing the operator by hand, given a model, whether the operator can be generated dynamically. The authors of Refs.\cite{rg7,rg8,rg9,rg10} showed that the Type-I seesaw mechanism which is otherwise studied for generating light neutrino masses and leptogenesis from RH neutrino decays\cite{lep1}, facilitates $\mathcal{L}_{CPV}$ at two loop-level (Fig.\ref{fig1}) even when the seesaw Lagrangian is minimally coupled to the gravitational background. Physically, a non-vanishing value of the chemical potential which could be attributed to an asymmetric propagation of lepton and anti-lepton can be understood by computing the  self energy diagrams for lepton and anti-lepton propagators (in gravitational background) which in the seesaw model leave different contributions  only at the two loop level (Fig.\ref{fig1}). 

To understand how $\mathcal{L}_{CPV}$ gets generated in seesaw model, one has to capitalise on the fact that the effective coupling `$b$' is independent of the choice of background and therefore the computation can simply be done in a conformally flat metric given by
\bea
g_{\mu\nu}=(1+h)\eta_{\mu\nu}\label{conf}
\eea
so that the contribution from $\mathcal{L}_{CPV}= b\partial_\mu R\bar{\ell}\gamma^\mu\ell$ to the effective $\ell\ell h$ vertex in the momentum space reads
\bea
A(q)=3ib(q^2\slashed{q})h(q),\label{mp}
\eea
where $q=p^\prime-p$ is the momentum transfer between the ingoing ($p^\prime$) and outgoing lepton($p$) and the Ricci scalar is given by $R=-3\partial^2h$. In the seesaw model, a construction  of an effective $\ell\ell h$ vertex that leaves a similar contribution as in Eq.\ref{mp} would then manifest the generation $\mathcal{L}_{CPV}$ operator. The coupling $`b$' therefore can be  calculated by matching the terms proportional to $q^2\slashed{q}$. Now looking at Fig.\ref{fig1}, it is clear that to create an effective $\ell\ell h$ vertex, one needs to have an insertion of $h$ for example by the means of $NNh$, $HHh$, $H\ell Nh$ etc. terms. Then the function $A(q)$ can be calculated by computing the transition matrix element $\braket{\ell_\alpha
(p^\prime)|\mathcal{O}h|\ell_\alpha(p)}$, where the operator $\mathcal{O}h$ can be generated using Eq.\ref{conf} and a proper conformal rescaling\cite{rg7,rg10}of the fields followed by an expansion (upto linear order in $h$)of  UV Lagrangian
\bea
\mathcal{L}=\mathcal{L}_{EW}+\sqrt{-g}\left[i\bar{N}_{Ri} \slashed{D}N_{Ri}-\left \lbrace f_{\alpha i}\bar{\ell}_{L\alpha}\tilde{H} N_{Ri}
+\frac{1}{2}\bar{N}_{Ri}^C(M_R)_{ij} \delta _{ij}N_{Rj} 
+ {\rm h.c.}\right\rbrace\right]\,, \label{seesawlag0}
\eea
where $\sqrt{-g}$ is the square root of the metric determinant, $\ell_{L\alpha}=\begin{pmatrix}\nu_{L\alpha} & e_{L\alpha}\end{pmatrix}^T$ is the SM lepton doublet of flavour $\alpha$, $\tilde{H}=i\sigma^2 H^*$ with $H= \begin{pmatrix}H^+&H^0
\end{pmatrix}^T $ being the Higgs doublet and $M_R={\rm diag}\hspace{.5mm} (M_1,M_2,M_3)$, $M_{1,2,3}>0$. It can be shown that only the $NNh$ insertion is what is relevant to the computation of the transition matrix element and essentially one has to compute four two-loop diagrams (two diagrams for each of the diagrams in Fig.\ref{fig1} with $N_iN_ih$ and $N_jN_jh$ insertion)\cite{rg8,rg10}.  The contribution from these diagrams to the matrix element reads as 
\bea
A(q)=i(q^2\slashed{q})h(q)\sum_{i,j,\beta}\frac{{\rm Im} \left[f_{i\alpha}^\dagger f_{\alpha j}f_{i\beta}^\dagger f_{\beta j}\right]}{M_{i}M_{j}}I_{[ij]},\label{maeel}
\eea
where the loop function $I_{[ij]}$ depends on the heavy neutrino masses. Comparing Eq.\ref{mp} and Eq.\ref{maeel}, it is evident that the effective coupling $b$ is simply given by 
\bea
b=\sum_{i,j,\beta}\frac{{\rm Im} \left[f_{i\alpha}^\dagger f_{\alpha j}f_{i\beta}^\dagger f_{\beta j}\right]}{3M_{i}M_{j}}I_{[ij]}
\eea
and consequently, $\mathcal{L}_{CPV}$ operator for one generation of leptons in seesaw model reads as
\bea
\mathcal{L}_{CPV}=\sum_{i,j,\beta}\frac{{\rm Im} \left[f_{i\alpha}^\dagger f_{\alpha j}f_{i\beta}^\dagger f_{\beta j}\right]}{3M_{i}M_{j}}I_{[ij]} \partial_\mu R\bar{\ell_\alpha}\gamma^\mu\ell_\alpha.
\eea
Now recalling  the chemical potential $\mu=b\dot{R}$, using standard Fermi-Dirac statistics for lepton and anti-lepton equilibrium densities and normalising the lepton asymmetry with photon number density, the net lepton asymmetry (summing over all the lepton generations) can be calculated as
\bea
N_{B-L}^{eq}=\frac{\pi^2\dot{R}}{36}\sum_{j>i}\frac{{\rm Im}\left[k_{ij}^2\right]}{\zeta (3)T M_i M_j}I_{[ij]},\label{gl1}
\eea
where $k_{ij}=(f^\dagger f)_{ij}$. The general form of the loop function $I_{[ij]}$ is given by 
\bea
I_{[ij]}=\frac{1}{(4\pi)^4}\left(\frac{M_j^2}{M_i^2}\right)^p{\rm ln }\left(\frac{M_j^2}{M_i^2}\right)
\eea
where $p=0,1$\cite{rg7,rg8,rg9,rg10}. Note that in Refs.\cite{rg7,rg8,rg9,rg10}, the authors show, though $p=0$ is a conservative solution, $p=1$  is also strongly preferred (see e.g., the  discussion related to `Diagram 4' in sec 4.1 of Ref.\cite{rg8}. Briefly, the transition amplitude can be shown to be a difference between the amplitudes of two self-energy diagrams, where the first one shows a clear $p=1$ behaviour and the second  one which requires a non-trivial analytic computation of ten scalar topologies, from the power counting argument can be shown to be not dominant enough to cancel the amplitude of the first diagram. However, still there could be a room for an unlikely conspiracy for some of the scalar topologies of the second diagram to cancel the $p=1$ behaviour of the first diagram. That fine-tuned region of parameter space requires full analytic computation of all the scalar topologies for the second diagram). Here we do not confront the robustness of $p=1$ solution (or the `hierarchically enhanced solution'\cite{rg9,rg10}), rather we take it at face value. As an aside, let us mention that  $p=0$ solution does not work (cannot produce correct baryon asymmetry) in the standard cosmological evolution of the universe\cite{rg9,rg10,Samanta:2020tcl}. From now on we shall proceed with hierarchically enhanced equilibrium asymmetry 
\bea
N_{B-L}^{eq}=\frac{\pi^2\dot{R}}{36 (4\pi )^4}\sum_{j>i}\frac{{\rm Im}\left[k_{ij}^2\right]}{\zeta (3)T M_i M_j}\left(\frac{M_j^2}{M_i^2}\right){\rm ln }\left(\frac{M_j^2}{M_i^2}\right),\label{gl1}
\eea
considering a standard cosmological evolution and therefore, we stress that any conclusive future demurral of $p=1$ would imply our analysis is invalid.
To proceed further, the time derivative of the Ricci scalar $\dot{R}$ is given by 
\bea
\dot{R}=\sqrt{3}\sigma^{3/2}(1-3 \omega)(1+ \omega) \frac{T^6}{M_{Pl}^3},
\eea
where $\sigma=\pi^2 g^*/30 $, $M_{Pl}\sim 2.4\times 10^{18}$  GeV and $\omega$ being the equation of state parameter.  A non-zero value of $\dot{R}$ in radiation domination is obtained in all the usual scenarios of gravitational leptogenesis by considering so called trace-anomaly in the gauge sector allowing $1-3\omega\simeq 0.1$\footnote{Note that $\delta \omega=1-3\omega\simeq 0.01-0.1\neq 0$ is a crucial ingredient of all the gravitational leptogenesis models. In the SM,  $\delta \omega\simeq 0$ is still a very good assumption at the higher temperatures\cite{trac}. Therefore, most of the  gravitational leptogenesis models that assume a large and nonvanishing $\delta\omega$, intrinsically refer to a BSM theory. For example, without going into the detail of a specific BSM model, starting from the Ref.\cite{tra} to the Refs.\cite{rg7,rg8,rg9,rg10} refer to $\delta\omega\sim  0.1$ considering the finite temperature QCD as $SU(N_c)$ gauge theory with $N_f$ massless quark flavours and coupling $g$ which give rise to $\delta\omega$ as\cite{Kajantie:2002wa} \bea
\delta\omega= \frac{5}{12\pi^2}\frac{g^4}{(4\pi)^2}\frac{\left(N_c+\frac{5}{4}N_f\right)\left(\frac{11}{3} N_c-\frac{2}{3}N_f\right)}{2+\frac{7}{2}\frac{N_cN_f}{N_c^2-1}}+\mathcal{O}(g^5)
\eea
so that typical gauge groups and matter content can easily yield $\delta\omega\simeq 0.01-0.1$ at high energy\cite{tra}. Note that we quote the result of Eq.3.88 of Ref.\cite{trac} which differs from the results of Eq.4 of Ref.\cite{tra} by a factor 2.}\cite{tra}. The light neutrino masses are obtained from the flat space seesaw Lagrangian and are given in the flavour basis  as
\bea
M_\nu = -m_DM_R^{-1}m_D^T\,. \label{seesaweq}
\eea
where $m_D=fv$ with $v=174$ GeV being the vacuum expectation value of the SM Higgs. Neutrino-less double beta decay parameter is the absolute value of the (11) element of $M_\nu$, i.e $|M_{\nu,11}|\equiv |m_{\beta\beta}|$\cite{Rodejohann:2011mu}. The mass matrix in Eq.\ref{seesaweq} can be diagonalised by a unitary matrix $U$ as
\bea
U^\dagger m_D M_R^{-1}m_D^T U^*=D_m,\label{see2}
\eea
where $D_m=-~{\rm diag}~(m_1,m_2,m_3)$ with $m_{1,2,3}$ being the physical light neutrino masses. We work in a basis where the RH neutrino mass matrix $M_R$ and charged lepton mass matrix $m_\ell$ are diagonal. Therefore, the neutrino mixing matrix $U$ can be written as 
\bea
U=P_\phi U_{PMNS}\equiv 
P_\phi \begin{pmatrix}
c_{1 2}c_{1 3} & s_{1 2}c_{1 3} & s_{1 3}e^{-i\delta }\\
-s_{1 2}c_{2 3}-c_{1 2}s_{2 3}s_{1 3} e^{i\delta }&  c_{1 2}c_{2 3}-s_{1 2}s_{1 3} s_{2 3} e^{i\delta} & c_{1 3}s_{2 3} \\
s_{1 2}s_{2 3}-c_{1 2}s_{1 3}c_{2 3}e^{i\delta} &-c_{1 2}s_{2 3}-s_{1 2}s_{1 3}c_{2 3}e^{i\delta} & c_{1 3}c_{2 3}
\end{pmatrix}P_M\,,\nonumber\\
\label{eu}
\eea
where $P_M={\rm diag}~(e^{i\alpha_M},~1,~e^{i\beta_M})$  is the Majorana phase matrix,  $P_\phi={\rm diag}~(e^{i\phi_1},~e^{i\phi_2},~e^{i\phi_3})$  is an unphysical diagonal phase matrix and $c_{ij}\equiv\cos\theta_{ij}$, $s_{ij}\equiv\sin\theta_{ij}$ with the mixing angles $\theta_{ij}=[0,\pi/2]$. Low energy CP violation enters in Eq.\,\ref{eu} via the Dirac phase $\delta$ and the Majorana phases $\alpha_M$ and $\beta_M$. It is useful to parametrise (which can be straightforwardly derived from Eq.\ref{see2}) the Dirac mass matrix as\cite{Casas:2001sr}
\bea
m_D=U\sqrt{D_m}\Omega\sqrt{M_R},\label{orth}
\eea
where $\Omega$ is a $3\times 3$ complex orthogonal matrix and is given by
 \bea
\Omega = 
\left( \begin{array}{ccc}
1 &  0 & 0 \\
0 & \cos z_{23} & \sin z_{23} \\
0 & - \sin z_{23}  & \cos z_{23}
\end{array}\right) \,  
\left( \begin{array}{ccc}
\cos z_{13} & 0  & \sin z_{13} \\
0 & 1 & 0 \\
- \sin z_{13} & 0 & \cos z_{13}
\end{array}\right) \,  
\left( \begin{array}{ccc}
\cos z_{12} & \sin z_{12} & 0 \\
-\sin z_{12} & \cos z_{12} & 0 \\
0 & 0 & 1
\end{array}\right)  \,  ,
\eea
where $z_{ij}=x_{ij}+i y_{ij}$. In the hierarchical limit of the RH neutrinos $M_3\gg M_2\gg M_1$, the equilibrium asymmetry can be approximated as
 \bea
N_{B-L}^{eq}\simeq \frac{\pi^2 \dot{R}}{36 (4\pi v)^4}\frac{\sum_{k,k^\prime} m_km_{k^\prime} {\rm Im\left[ \Omega_{k1}^*\Omega_{k3}\Omega_{k^\prime 1}^*\Omega_{k^\prime 3}\right]}}{\xi(3) T} \frac{M_3^2}{M_1^2}{\rm ln \left(\frac{M_3^2}{M_1^2}\right)}.\label{nblg0}
\eea

As mentioned previously, one needs to compute the frozen out asymmetry $N_{B-L}^{G0}$ considering the effect of $\Delta L=2$ processes which tend to maintain the asymmetry $N_{B-L}^{eq}$ in equilibrium and therefore a dilution of the asymmetry from $z_{in}$ upto $z_0$-the freeze-out point of the asymmetry. The  asymmetry $N_{B-L}^{G0}$ can be obtained by solving a simple Boltzmann equation\cite{rg10}
\begin{figure}
\includegraphics[scale=.4]{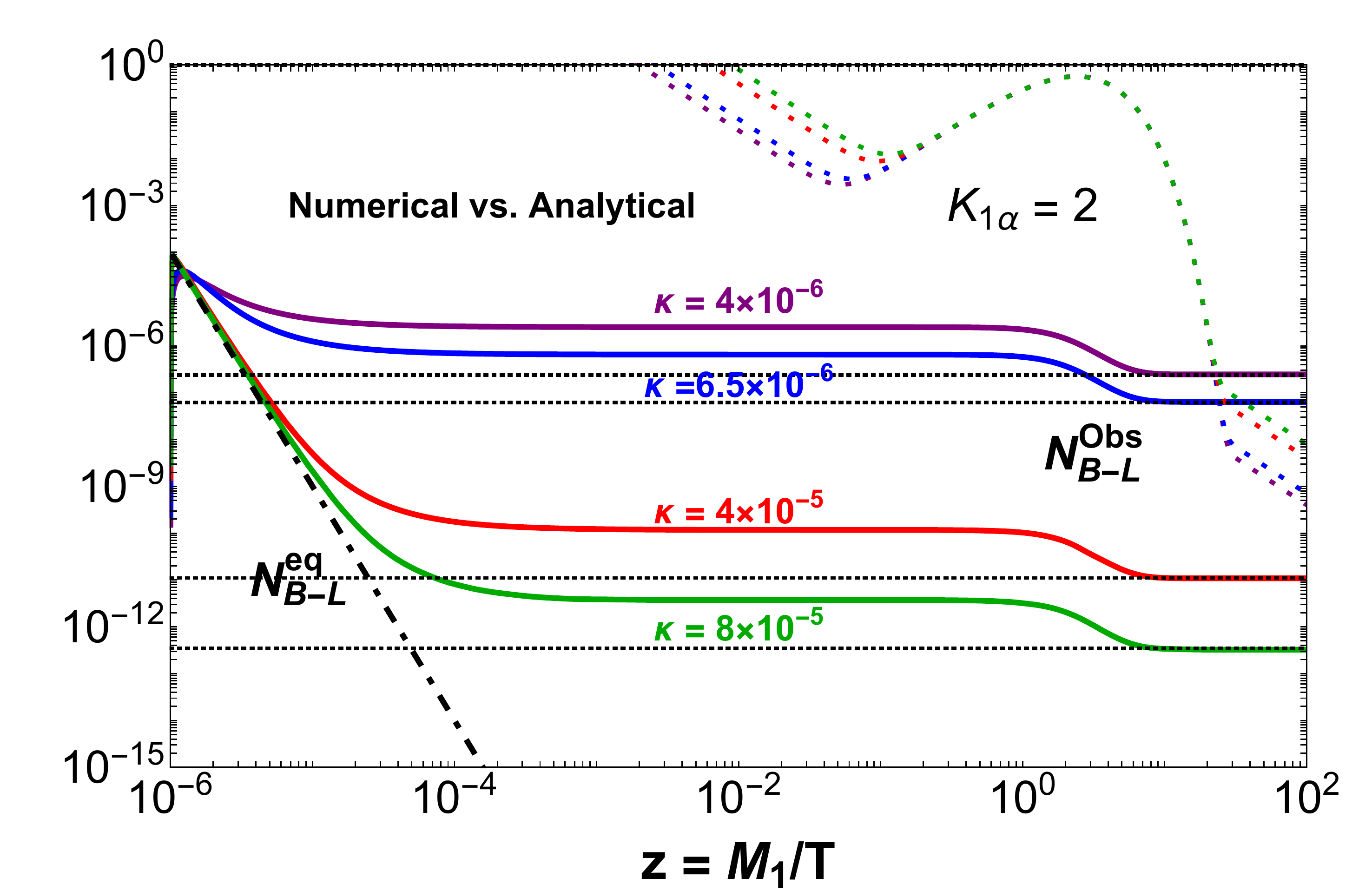}
\caption{A numerical vs. analytical comparison of $N_{B-L}^{G0}$. The coloured solid lines are numerical solutions and the black dashed lines are analytical yields. The coloured dashed lines are the lepton number violating interactions--$\Delta L=2+$Inverse decays which contribute to the washout processes (initial dilution and a second stage washout).}\label{nva}
\end{figure}
\bea
\frac{dN_{B-L}}{dz}=-W_{\Delta L=2}(N_{B-L}-N_{B-L}^{eq}),\label{BE}
\eea
where $W_{\Delta L=2}$ encodes the effect of $\Delta L= 2$ process  involving non-resonant $N_1$-exchange and is given by \cite{pre1,rg10}
\bea
W_{\Delta L=2}(z\ll 1)\simeq \frac{12m^* M_1}{\pi^2 v^2 z^2}\left( \left[\frac{\bar{m}}{m^*}\right]^2+K_1^2-\frac{2m_1^2}{m^{*2}}\right)~~{\rm with}~~\bar{m}=\sqrt{\sum_i m_i^2}.
\eea
For a parametric scan using $3\sigma$ neutrino oscillation data\cite{globalfit}, it is convenient to solve the BE in Eq.\ref{BE} analytically. To this end, we re-write Eq.\ref{BE} as
\bea
\frac{dN_{B-L}}{dz}=-\frac{\kappa}{z^2}\left(N_{B-L}-\frac{\beta}{z^5}\right),
\eea
where 
\bea
\kappa=\frac{12m^* M_1}{\pi^2 v^2 }\left( \left[\frac{\bar{m}}{m^*}\right]^2+K_1^2-\frac{2m_1^2}{m^{*2}}\right),~\beta=\frac{\sqrt{3}\sigma^{3/2}M_1^5}{M_{Pl}^3}(1-3\omega)(1+\omega)\mathcal{Y}
\eea
and the unflavoured $N_1$-decay parameter $K_1$ is given in terms of orthogonal matrix as 
\bea
K_i=\frac{1}{m^*}\sum_k m_k |\Omega_{ki}|^2.\label{totdecay}
\eea 
The parameter $\mathcal{Y}$ which encodes the CP violation in the theory is given by
\bea
\mathcal{Y}= \frac{\pi^2}{36 (4\pi v)^4}\frac{\sum_{k,k^\prime} m_km_{k^\prime} {\rm Im\left[ \Omega_{k1}^*\Omega_{k3}\Omega_{k^\prime 1}^*\Omega_{k^\prime 3}\right]}}{\xi(3)} \frac{M_3^2}{M_1^2}{\rm ln \left(\frac{M_3^2}{M_1^2}\right)}.
\eea 
Starting from a vanishing initial abundance of $N_{B-L}(z)$, for large values of $z$ one finds the analytical solution for $N_{B-L}^{G0}$ as
\bea
N_{B-L}^{G0}=\frac{120\beta}{\kappa^5}\left[1-e^{-\kappa/z_{\rm in}}\right]-\frac{\beta e^{-\kappa/z_{\rm in}}}{\kappa^5}\left[\sum_{n=1}^5\frac{5!}{n!}\left(\frac{\kappa}{z_{\rm in}}\right)^n\right].\label{key}
\eea
Since the lightest RH scale is below $10^9$ GeV, a second stage $N_1$-washout by inverse decays occurs in all the three flavours and therefore the final asymmetry is given by\cite{Samanta:2020tcl}
\bea
N_{B-L}^f=\sum_{\alpha = e, \mu, \tau}N_{\Delta \alpha} =\frac{1}{3}\sum_{\alpha = e, \mu, \tau}\left(\frac{120\beta}{\kappa^5}\left[1-e^{-\kappa/z_{\rm in}}\right]-\frac{\beta e^{-\kappa/z_{\rm in}}}{\kappa^5}\left[\sum_{n=1}^5\frac{5!}{n!}\left(\frac{\kappa}{z_{\rm in}}\right)^n\right]\right)e^{-\frac{3\pi}{8}K_{1\alpha}},\label{key}\nonumber\\
\eea
where the flavoured washout parameters are given by 
\bea
K_{i\alpha}=\frac{1}{m^*}\left | \sum_k U_{\alpha k}\sqrt{m_k}\Omega_{ki}\right |^2.\label{fdp}
\eea
One has to compare Eq.\ref{key} with the measured asymmetry at recombination 
\bea
\eta_{CMB}\simeq 10^{-2}N_{B-L}^f\simeq 6\times 10^{-10}.
\eea
In Fig.\ref{nva}, we show the dynamics of the lepton asymmetry production (inclusive of a second stage $N_1$-washout for benchmark value of $K_{1\alpha}=2$). As one sees as $\kappa$ increases $\Delta L=2$ interactions try to maintain the asymmetry in equilibrium for a longer period of time and hence causes a late freeze-out as well as a reduction in magnitude of the frozen out asymmetry. Thus, if the elements of the orthogonal matrix are not significantly large\cite{DiBari:2014eqa}-which also correspond to a fine tuning in the seesaw formula\cite{DiBari:2018fvo,brome}, an increase of $m_1$ causes an increase of the value of $\kappa$ (shown in Fig.\ref{fig3} ) and consequently the magnitude of the asymmetry reduces. This leads to an upper bound on $m_1$ for successful gravitational leptogenesis. Note that in the analytical formula for $N_{B-L}^f$ we have neglected the flavour effect in the $\Delta L=2$ process as well as  all the possible values of the flavour projectors (that project the asymmetry on the $e,\mu,\tau$ basis)  compatible with $3\sigma$ oscillation data. In Ref.\cite{Samanta:2020tcl} we computed it numerically and found these effects do not have any significant effect on the final asymmetry. For convenience, let us re-mention here that the frozen out asymmetry that is large enough in magnitude to be compatible with the observed one, corresponds to smaller values $\kappa$ which are not very sensitive to the flavour effect. On the other hand, in the probability triangle, though  flavour projectors show a biasness towards the electron flavour, $P_{i\alpha}=1/3$ still remains a fair choice to take under consideration\cite{DiBari:2018fvo,Samanta:2020tcl}. Therefore to efficiently scan the entire parameter space in the computer codes, we have simplified the final formula for $N_{B-L}^f$ considering a democratic behaviour of the flavour projectors. A consideration of  the full $3\sigma$ data for the flavour projectors (analytical formula including flavour projectors is given in Ref.\cite{Samanta:2020tcl}) would affect the final results only at the level of few percent. However, let us mention that in our previous analysis we did not take into account the effect of flavour-couplings (FC) at the $N_1$-washout\cite{fl2,Samanta:2018efa,fc1,fc2,fc3}. Though in typical leptogenesis studies, FCs are included for more accurate computation involving flavour effect, in some scenarios, FCs play roles which are of great interest\cite{Samanta:2018efa,fc3}.  For example, in Ref.\cite{Samanta:2018efa} it has been shown that even if CP violation is absent in one particular flavour, FCs can generate significant lepton asymmetry in that flavour.  In this article we include the effect of FCs towards a fuller treatment of flavour effect and in the numerical section we shall state the percentage of correction to the final result. The flavour coupling effect is introduced in the washout equation as 
\bea
\frac{dN_{\Delta_\alpha}}{dz}=-P_{1\alpha}C_{\alpha \beta}W_1^{\rm ID}N_{\Delta_\beta},\label{flcc}
\eea
where $P_{1\alpha}=K_{1\alpha}/K_1$ and the flavour coupling matrices in the three flavour regime\cite{fc2,Samanta:2018efa}
\bea
C_{l}=\begin{pmatrix}
151/179&-20/179&-20/179\\
-25/358&344/537&-14/537\\
-25/358&-14/537&344/537
\end{pmatrix}, C_{h}=\begin{pmatrix}
37/179&52/179&52/179\\37/179&52/179&52/179\\37/179&52/179&52/179
\end{pmatrix}, 
\eea
yield
\bea
C_{\alpha\beta}\equiv C_{l}+C_{h}=\begin{pmatrix}
188/179&32/179&32/179\\49/358&500/537&142/537\\49/358&142/537&500/537
\end{pmatrix}.
\eea
Clearly, unlike Eq.\ref{key} which is a solution of Eq.\ref{flcc} with $C=\mathbb{I}$, now the equation for the final asymmetry would be more complicated and interestingly, a particular flavour component of the asymmetry would receive contribution from other flavours. For computational purpose, it is convenient to perform a basis rotation to make the Eq.\ref{flcc} diagonal in a redefined flavour basis.  To this end, Eq.\ref{flcc} can be written in matrix form as
\bea
\frac{d\vec{N}_{\Delta}}{dz}=-W_1^{\rm ID}\tilde{P}_1\vec{N}_{\Delta},\label{flcc1}
\eea
where $\vec{N}_{\Delta}=\left(N_{\Delta e},N_{\Delta\mu},N_{\Delta\tau}\right)^T$ and $\tilde{P}_1=P_{1\alpha}C_{\alpha \beta}$. Introducing the $V$ matrix that diagonalises $\tilde{P}_1$ as $V\tilde{P}_1V^{-1}=\tilde{P}_1^\prime$, Eq.\ref{flcc1} can be written in the new flavour basis as
\bea
\frac{d\vec{N}^\prime_{\Delta}}{dz}=-W_1^{\rm ID}\tilde{P}^\prime_1\vec{N}^\prime_{\Delta},\label{flcc2}
\eea
where $\vec{N}^\prime_{\Delta}=V\vec{N}_{\Delta}$. Therefore, the asymmetry matrix in the prime is simply obtained as
\bea
\vec{N}^{\prime f}_{\Delta}=\left(N_{\Delta e^\prime}^{G0}e^{-\frac{3\pi}{8} K_{1e^\prime}}~~N_{\Delta \mu^\prime}^{G0}e^{-\frac{3\pi}{8} K_{1\mu^\prime}}~~N_{\Delta \tau^\prime}^{G0}e^{-\frac{3\pi}{8} K_{1\tau^\prime}}\right)^T,
\eea
where $K_{1\alpha^\prime}=P_{1\alpha^\prime} K_1$.
Consequently, the final asymmetry matrix which we meed in the unprimed basis is obtained as
\bea
\vec{N}_{\Delta}^f=V^{-1}\left(\sum_{\beta}V_{e^\prime\beta}N_{\Delta \beta}^{G0}e^{-\frac{3\pi}{8} K_{1e^\prime}}~~\sum_{\beta}V_{\mu^\prime\beta}N_{\Delta \beta}^{G0}e^{-\frac{3\pi}{8} K_{1\mu^\prime}}~~\sum_{\beta}V_{\tau^\prime\beta}N_{\Delta \beta}^{G0}e^{-\frac{3\pi}{8} K_{1\tau^\prime}}\right)^T.
\eea 
Therefore, the total asymmetry that produces the observed baryon asymmetry is given by
\bea
N_{B-L}^f=\sum_\alpha N^f_{\Delta \alpha}=\sum_\alpha\sum_{\alpha^\prime} V^{-1}_{\alpha \alpha^\prime}\sum_{\beta}V_{\alpha^\prime\beta}N_{\Delta \beta}^{G0}e^{-\frac{3\pi}{8} K_{1\alpha^\prime}}.\label{flvcor}
\eea
\section{Gravitational waves from cosmic string}\label{s3}
Cosmic strings are natural prediction of many extension of standard model featuring $U(1)$ symmetry breaking\cite{cs2,Nielsen:1973cs}. They are considered to be one dimensional object having string tension $\mu$ which is typically taken to be of the order of the square of the symmetry breaking scale. The normalised string tension $G\mu$, with $G$ being the Newton's constant, is directly constrained by CMB  as $G\mu\lesssim  1.1\times 10^{-7}$\cite{stcmb}. After formation, the strings are expected to  reach a scaling regime in which their net energy density tracks the total  energy density of the universe with a relative fraction $G\mu$.  This regime is considered to have many closed loops and Hubble-length long strings which intersect to form new loops as the universe expands. All these loops oscillate and emit radiation, including gravitational waves.  We consider stochastic gravitational background  (SGWB) from cosmic string scaling by considering Nambu-goto strings which radiate energy dominantly in the form of GW radiation. We follow Ref.\cite{gr2} to calculate SGWB from cosmic string. Once the loops  are formed, they radiate energy in the form of gravitational radiation at a constant rate, mathematically described as
\bea
\frac{dE}{dt}=-\Gamma G\mu^2,
\eea
where $G$ is the usual gravitational constant and $\Gamma=50$\cite{Vachaspati:1984gt,Vilenkin:1981bx}. Thus, the initial length $l_i=\alpha t_i$ of the loop decreases as 
\bea
l(t)=\alpha t_i-\Gamma G\mu(t-t_i)
\eea
until the loop disappears completely. The quantity $\alpha$ has a distribution and for the largest loop one typically has $\alpha=0.1$\cite{Blanco-Pillado:2017oxo,Blanco-Pillado:2013qja} which we consider in the numerical calculation. The total energy loss from a loop is decomposed
into a set of normal-mode oscillations at frequencies $\tilde{f}=2k/l$, where $k=1,2,3..$. The relic GW density parameter is given by
\bea
\Omega_{GW}=\frac{f}{\rho_c}\frac{d\rho_{GW}}{df},
\eea
where $f$ is the red-shifted frequency and $\rho_c=3H_0^2/8\pi G$. The GW density  parameter $\Omega_{GW}$ can be written as a sum over all relic densities corresponding to a mode $k$ as
\bea
\Omega_{GW}(f)=\sum_k \Omega_{GW}^{(k)}(f),
\eea
where 
\bea
\Omega_{GW}^{(k)}(f)=\frac{1}{\rho_c}\frac{2k}{f}\frac{\mathcal{F}_\alpha \Gamma^{(k)} G\mu^2}{\alpha(\alpha+\Gamma G\mu)}\int_{t_F}^{t_0}d\tilde{t}\frac{C_{\rm eff}(t_i^{(k)})}{t_i^{(k)4}}\left[\frac{a(\tilde{t})}{a(t_0)}\right]^5\left[\frac{a(t_i^{k})}{a(\tilde{t})}\right]^3\Theta(t_i^{(k)}-t_F)
\eea
and the integration runs over the emission time with $t_F$ as time corresponding to the scaling regime of the loop after formation. The numerical values of $C_{\rm eff}$ are found to be 5.7 and 0.5 at radiation and matter domination and $\mathcal{F}_\alpha$ has a value $\sim 0.1$\cite{Blanco-Pillado:2017oxo,Blanco-Pillado:2013qja}. The quantity $t_i^{(k)}$ is the formation time of the loops contributing to the mode $k$ and is given by
\bea
t_{i}^{(k)}(\tilde{t},f)=\frac{1}{\alpha+\Gamma G \mu}\left[\frac{2k}{f}\frac{a(\tilde{t})}{a(t_0)}+\Gamma G \mu \tilde{t}\right].
\eea
The relative emission rate per mode is given by
\bea
\Gamma^{(k)}=\frac{\Gamma k^{-4/3}}{\sum_{m=1}^\infty m^{-4/3}}
\eea
with $\sum_{m=1}^\infty m^{-4/3}\simeq 3.6$ and $\sum_k\Gamma^{(k)}=\Gamma$. Having set up all the theoretical machineries, we now proceed towards the final discussion containing numerical results.
\section{Numerical results and discussions}\label{s4}
\begin{figure}
\includegraphics[scale=.5]{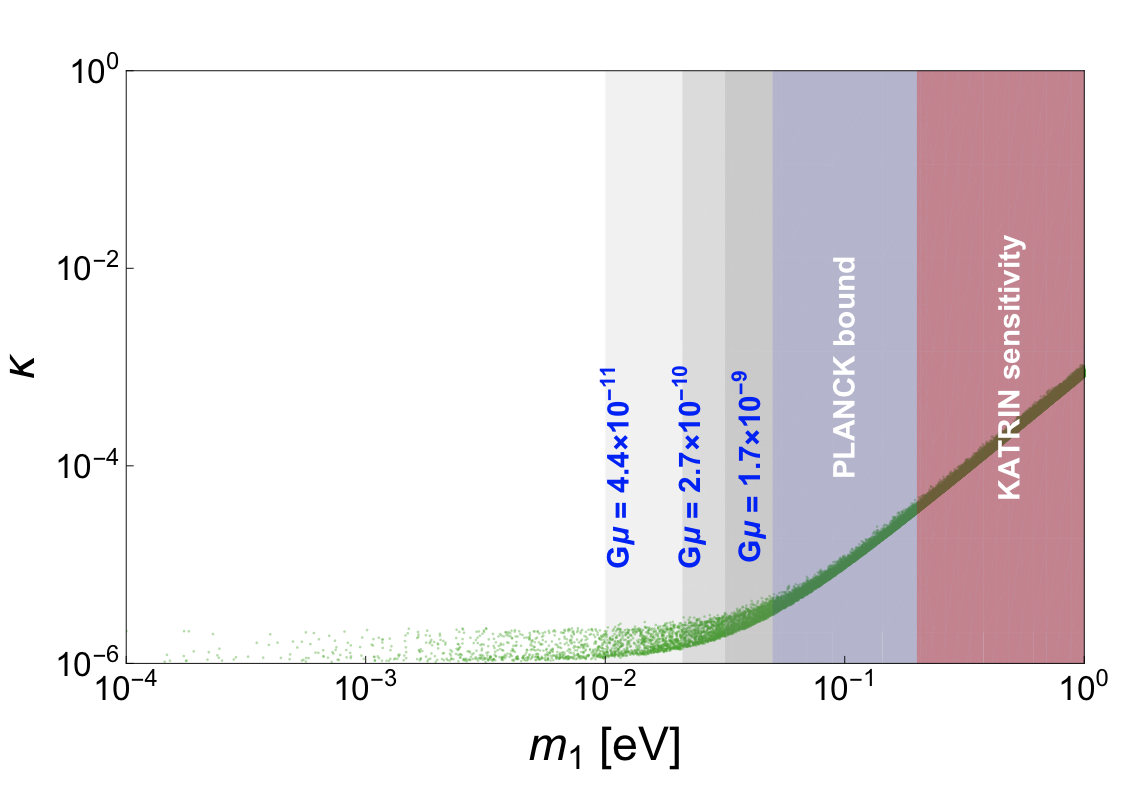}\\
\includegraphics[scale=.5]{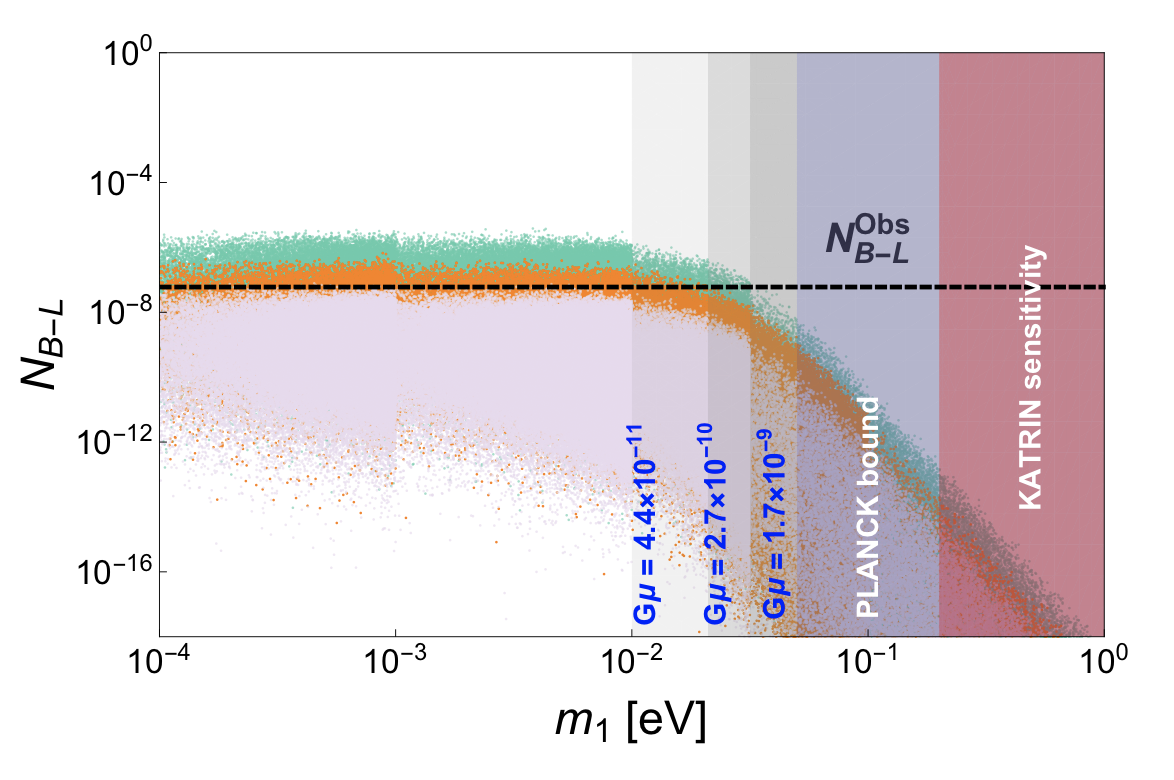}\\
\hspace{5mm}\includegraphics[scale=.5]{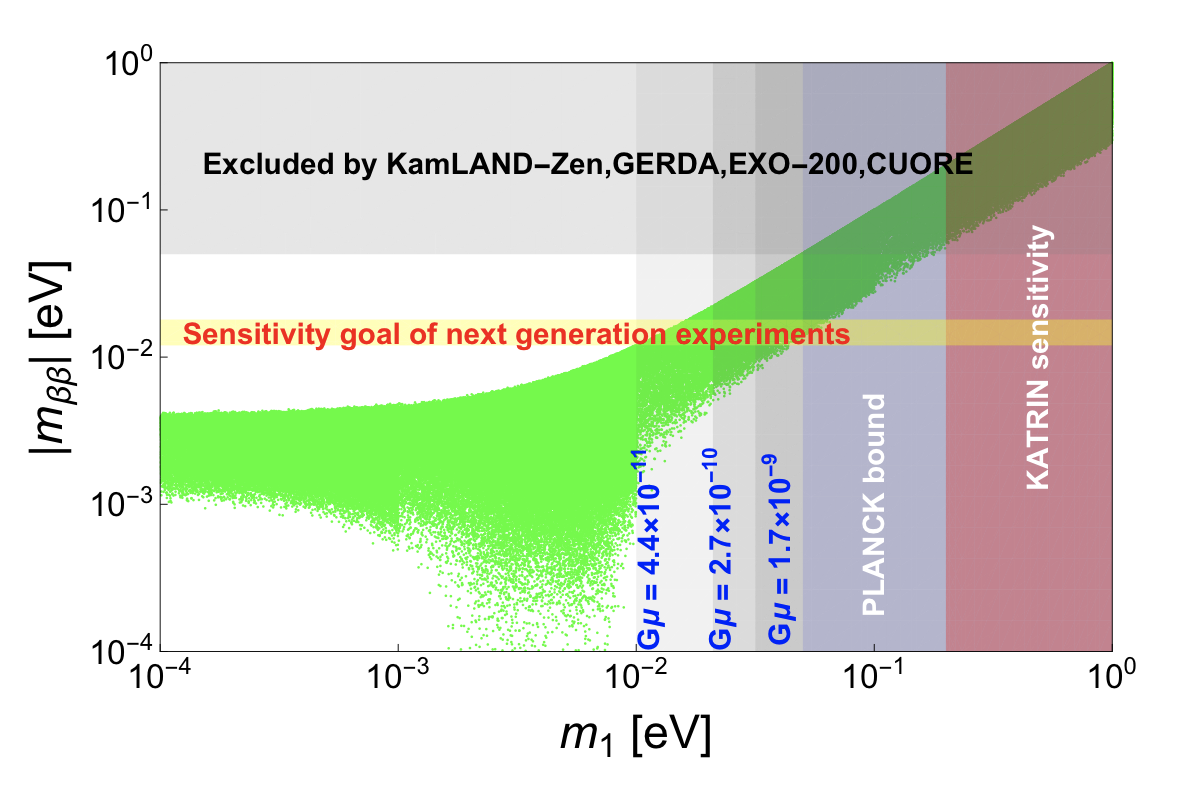}
\caption{Upper panel: $m_1$ vs. $\kappa$. Middle panel $m_1$ vs. $N_{B-L}^f$. Bottom panel: $m_1$ vs. $|m_{\beta\beta}|$.}\label{fig3}
\end{figure}

To generate all the plots in Fig.\ref{fig3}, we scanned over 3$\sigma$ neutrino oscillation data\cite{globalfit} and used the seesaw fine-tuning parameter $\gamma_i=\sum_j |\Omega^2_{ij}|\simeq 1$ which also helps to avoid the non-perturbative Yukawa couplings, i.e., ${\rm Tr}(f^\dagger f)\leq 4\pi$. We use the upper bound on the sum of the light neutrino masses as $\sum_i m_i<0.17$ eV\cite{planck} which corresponds to $m_1\lesssim 50$ meV as shown by vertical light blue shade in each of the plots. A more stringent upper bound $m_1\lesssim 31$ meV is also available from latest PLANCK data\cite{Akrami:2018vks,Vagnozzi:2017ovm}. The red vertical region is the future sensitivity region of the KATRIN experiment which is starting to measure neutrino masses with an ultimate sensitivity to 0.2 eV\cite{Giuliani:2019uno}. In the top panel of the figure, we show the variation of $\kappa$ with $m_1$ which indicates that for $m_1\gtrsim 10^{-2}$ eV, $\kappa$ increases rapidly. An immediate consequence can be seen in the middle panel where $N_{B-L}$ has a decreasing slope for $m_1\gtrsim 10^{-2}$ eV. This corresponds to the previously mentioned late freeze out solutions as also shown in Fig.\ref{nva}. We show three gray shaded exclusion regions for the string tensions $G\mu=4.44\times 10^{-11},2.7\times 10^{-10}$ and $1.7\times 10^{-9}$ which correspond to the upper bounds $m_1\lesssim 10$ meV, 21 meV and 31 meV for successful leptogenesis. Corresponding exclusion regions on the effective matrix element of $0\nu\beta\beta$ decay have been shown in the bottom panel. The horizontal gray shaded region represents the already excluded region and the yellow shaded region represents the future sensitivity limits of next-generation $0\nu\beta\beta$ experiments. 
A comprehensive discussion about all the current and planned $0\nu\beta\beta$ experiments can be found in Ref.\cite{Giuliani:2019uno}. The above numerical discussion excludes the contribution of the FCs since it is sufficient to consider Eq.\ref{key} to have an overall idea of the parameter space. However, to obtain more  accurate upper bounds, it is instructive to include FCs as discussed in sec.\ref{s2}. Using Eq.\ref{flvcor} with a democratic behaviour of the flavour projectors, we perform a full numerical scan of the parameter space and find  bit more relaxed upper bounds on $m_1$. For the mentioned values of $G\mu$, we find $m_1\lesssim 12$ meV, 25 meV and 36 meV for successful leptogenesis. This implies in this scenario, FCs give correction around $17\%-20\%$ to the final result. We would like to take this opportunity to mention that we expect some level of  correction to the parameter space (which does not include FCs) in our previous publication\cite{Samanta:2020tcl} as well, where we discuss the flavour effects in RIGL mainly focusing on a two-RH neutrino scenario. A complete discussion in this context  will be presented elsewhere.\\

\begin{figure}
\includegraphics[scale=.45]{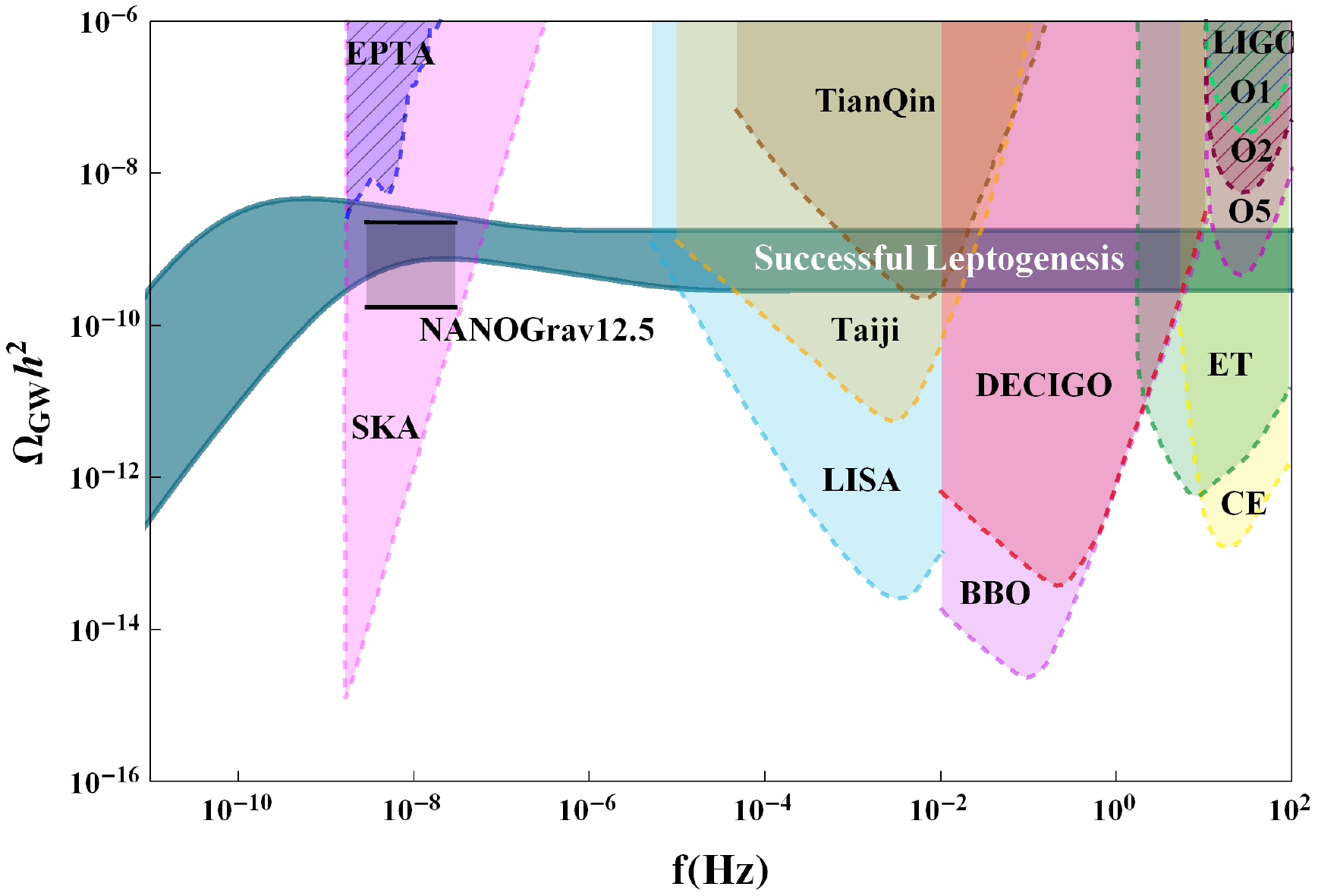}\includegraphics[scale=.4]{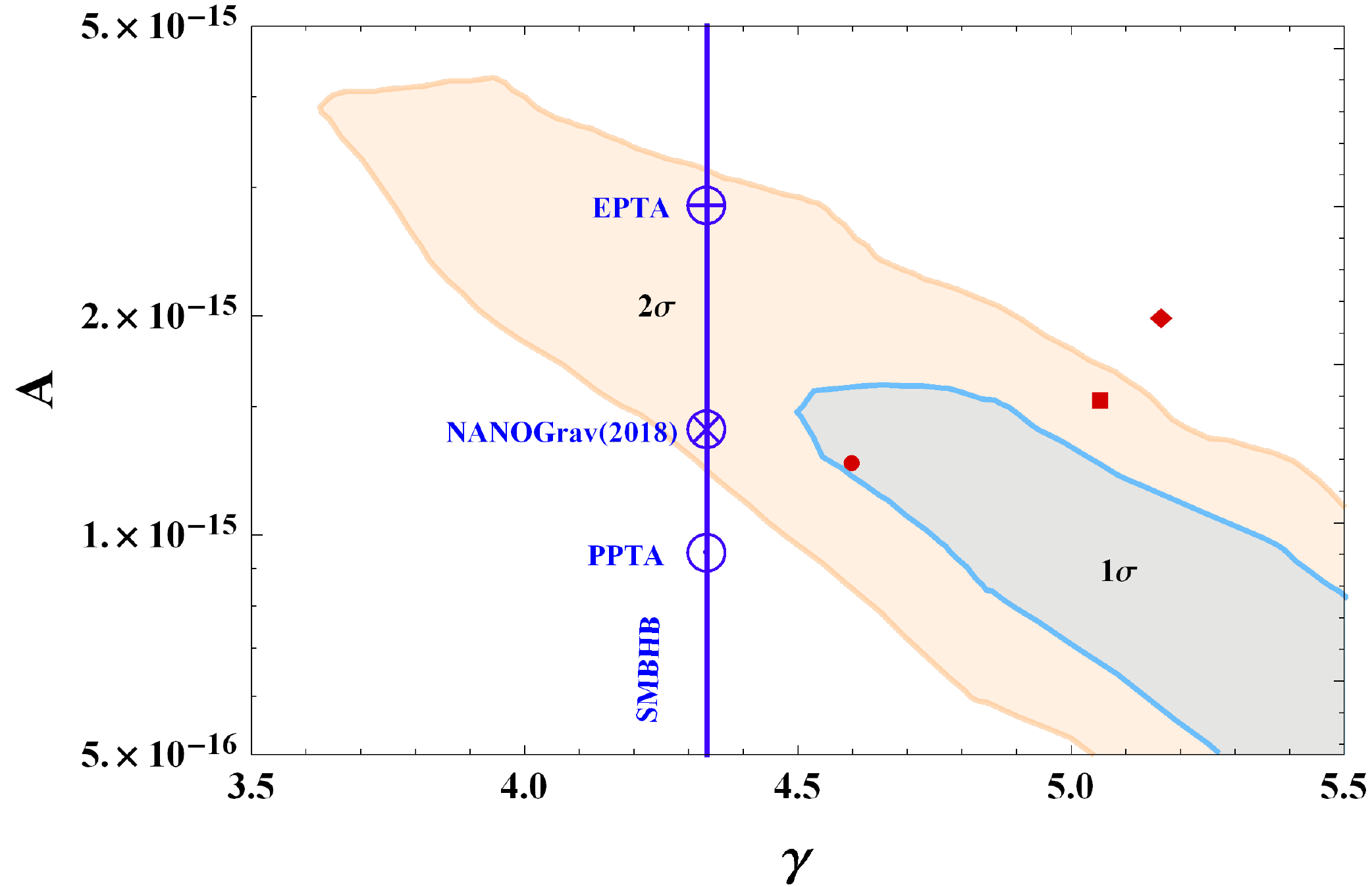}
\caption{Left: GW spectrum for sucessful RIGL mechanism. Right: Cosmic string fit to  NANOGrav data: $G\mu=4.44\times 10^{-11}$ (red circle), $2.7\times 10^{-10}$ (red squre) and $1.7\times 10^{-9}$ (red diamond).}\label{fig4}
\end{figure}

As one notices on the middle panel, a decrease in the string tension results in a decrease in the magnitude of the overall asymmetry which goes below the observed value ($N_{B-L}^{\rm Obs})$ for $G\mu<4.4\times 10^{-11}$. Therefore, RIGL will be fully tested in  the space-based  interferometers such as  LISA\cite{lisa}, Taiji \cite{taiji}, TianQin\cite{tianquin}, BBO\cite{bbo}, DECIGO\cite{decigo}, ground based interferometers like Einstein Telescope (ET)\cite{et} and Cosmic Explorer (CE)\cite{ce}, and atomic interferometers MAGIS\cite{magis}, AEDGE\cite{aedge} over  wide range of frequencies. Of course, as already argued, any exclusion of the string tension value $G\mu >4.4\times 10^{-11}$ by the GW detectors would put an exclusion region in the $|m_{\beta\beta}|$ parameter space or future discovery of $0\nu\beta\beta$ signal for $m_1 >10$ meV would put a lower bound on the string tension that would be tested by the GW detectors. In the left panel of Fig.\ref{fig4}, we show the GW spectrum that corresponds to an upper bound on $m_1$ within the range 12 meV-36 meV (bottom-up). We now conclude the paper by analysing the recent NANOGrav pulsar timing array (PTA) data  which if interpreted as GW signal, would put an interesting constraint on RIGL mechanism. \\ 

Though the idea of detecting GW with PTA is very much well known, for completeness we write few sentences. Pulsars are highly magnetised and rapidly rotating neutron stars. They emit radio waves from their magnetic poles and we observe these waves on the Earth as a string of pulses. Since neutron stars are of high densities, the time of arrival (TOA) of pulses are highly regular and that is why they are used in high precision timing experiments. Millisecond pulsars (spins $\sim$ 100 times a second) produce most stable pulses and are used by the PTAs. When a ``disturbance" like gravitational wave passes through the earth and pulsar system, the time of arrival of the  signal from the pulsars changes. This induces a frequency change in the pulses (contributes to a measurable quantity called time residual $R\propto \frac{\delta\nu}{\nu}$). The NANOGrav collaboration with their recently released data reports a strong evidence for a stochastic common-spectrum process (they analysed 45 pulsars) over independent-red noises\cite{nanog}.  However, they do not claim the detection as GW, since the time residuals do not show characteristic spatial relation described by the  Hellings–Downs (HD) curve\cite{Hellings:1983fr}.  In addition, other systematics such as pulsar spin noise\cite{spin} and solar system effects\cite{solar} might affect the signal thus the analysis requires proper handling of these two effects--a  study which is in preparation\cite{nanog}. In any case, if in the near future,  more data and a more rigorous statistical analysis by NANOGrav leads to the detection to SGWB, it would undoubtedly open up  a new direction to probe Early Universe cosmology. This of course includes leptogenesis as well. Remarkably enough, testing leptogenesis with pulsars would be a completely novel aspect  which can serve  also as a complementary probe of leptogenesis alongside the experiments in the particle physics side  such as neutrino oscillation  and neutrino-less double beta decay\cite{so1,so5}. Let us now focus on the analysis of gravitational leptogenesis scenario with respect to the NANOGrav data. 
The 12.5 yrs NANOGrav data are expressed in terms of power-law signal with characteristic strain given by 
\bea
h_c(f)=A\left(\frac{f}{f_{yr}}\right)^{(3-\gamma)/2}
\eea
with $f_{yr}=1 yr^{-1}$ and $A $ being the characteristic strain amplitude. The abundance of GWs has the standard form and can be recast as:
\bea
\Omega(f)=\Omega_{yr}\left(\frac{f}{f_{yr}}\right)^{5-\gamma}, ~~~~~{\rm with}~~~~~ \Omega_{yr}=\frac{2\pi^2}{3H_0^2}A^2f^2_{yr}.\label{po}
\eea
We do a simple power law fit to the cosmic string generated GW spectra using Eq.\ref{po} and show the results in the right panel of Fig.\ref{fig4} on the spectral index ($\gamma
$)-amplitude ($A$) plane against the NANOGrav@$1\sigma$ and $2\sigma$ contours. We plot the same benchmark values of $G\mu $ that were used in Fig.\ref{fig3}, i.e., $G\mu=4.44\times 10^{-11},2.7\times 10^{-10}$ and $1.7\times 10^{-9}$. These values are plotted as solid red circle, square and diamond points.  We find $G\mu=2.7\times 10^{-10}$ is at the edge of the $2\sigma$. Thus  RIGL disfavours $m_1>25$ meV at NANOGrav@2$\sigma$. Let's point out that our fit is consistent with Ref.\cite{nan1}, e.g., $G\mu\sim 2.7\times 10^{-10}$ is disfavored at 2$\sigma$, however, we get the strain amplitude value slightly lower than Ref.\cite{nan1}. The new NANOGrav 12.5 yr data\cite{nanog} though consistent with previous EPTA data\cite{Lentati:2015qwp}, they are in  tension with previous limits from PPTA \cite{Shannon:2015ect} and a previous NANOGrav analysis of their 11 yr data\cite{Arzoumanian:2018saf}. This tension would be reduced using improved  prior to the intrinsic pulsar red noise to the older data\cite{nanog}.
\section{summary}\label{s5}
We analyse a cosmic string induced GW spectrum as a test of leptogenesis. We discuss gravitational leptogenesis within the Type-1 seesaw which is otherwise studied in general for leptogenesis from right handed neutrino decays. An operator of the form $\mathcal{L}_{CPV}\sim b\partial_\mu R j^\mu\sim b\partial_\mu R\bar{\ell}\gamma^\mu\ell$ can generate a lepton asymmetry $N_{B-L}\sim \frac{b\dot{R}}{T}$ in thermal equilibrium evading Sakharov's third condition for baryogenesis. In seesaw model $\mathcal{L}_{CPV}$ can be created at two loop level with the right handed neutrinos as virtual particles. The generated equilibrium asymmetry is maintained (decreases with temperature) until the non-resonant  $\Delta L=2$ $N_1$-interaction goes out of equilibrium (then the asymmetry freezes out). The magnitude of the final asymmetry which depends on the lightest light neutrino mass $m_1$,  typically decreases with the $m_1$ and therefore the parameter space of the leptogenesis is sensitive to $m_1$ as well as the neutrino less double beta decay parameter $|m_{11}|$, through $m_1$. We consider that  the masses of the right handed neutrinos are generated dynamically by an $U(1)_{B-L}$ symmetry breaking which also leads to a formation of cosmic string network that produce gravitational waves. Therefore, the mechanism is sensitive to gravitational wave physics as well as low energy neutrino physics. We show that right handed neutrino induced gravitational leptogenesis can be probed  by the gravitational wave detectors as well as next-generation neutrinoless double beta decay experiments in a complementary manner such that  an exclusion limit on  $f-\Omega_{\rm GW}h^2$ plane would correspond to an exclusion on the $|m_{\beta\beta}|-m_1$ plane as well. We consider a normal light neutrino mass ordering and show that the gravitational wave detectors can fully test the mechanism for a wide range of frequencies. We then show that recent NANOGrav pulsar timing data (if interpreted as GW signal) would exclude $0\nu\beta\beta$ parameter space for $m_1\gtrsim 25$ meV at $\sim 2\sigma$.  
\section*{Acknowledgement}
The authors would like to thank Graham M. Shore for very helpful discussion about RIGL mechanism. This work was supported by  Newton International Fellowship (NIF 171202). This work was  supported also by European Structural and Investment Fund and the Czech Ministry of Education, Youth and Sports (Project International mobility MSCA-IF IV FZU - CZ.02.2.69/0.0/0.0/$20\_079$/0017754).
{}

\end{document}